# Holograms for power-efficient excitation of optical surface waves

Anton I. Ignatov,[1,2,3,*] Alexander M. Merzlikin,[1,2,3]

[1]*All-Russia Research Institute of Automatics, 22 ul. Sushchevskaya, Moscow 127055, Russia*
[2]*Moscow Institute of Physics and Technology, 9 Institutskiy per., Dolgoprudny, Moscow Region, 141700, Russia*
[3]*Institute for Theoretical and Applied Electromagnetics RAS, 13 Izhorskaya ul., Moscow 125412, Russia*
*\*Corresponding author: ignatovtoha@gmail.com*

**A method for effective excitation of optical surface waves based on holography principles has been proposed. For a particular example of excitation of a plasmonic wave in a dielectric layer on metal the efficiency of proposed volume holograms in the dielectric layer has been analyzed in comparison with optimized periodic gratings in the dielectric layer. Conditions when the holograms are considerably more efficient than the gratings have been found out. In addition, holograms recorded in two iterations have been proposed and studied. Such holograms are substantially more efficient than the optimized periodic gratings for all incidence angles of an exciting Gaussian beam. The proposed method is universal: it can be extended for efficient excitation of different types of optical surface waves and optical waveguide modes.**

## 1. INTRODUCTION

Surface-wave and plasmonic-wave optics are the actively developing fields of modern optics [1]. They are widely applicable for Raman spectroscopy [2,3], optical chemical and biosensors [4], optical interconnects [5,6] and modulators [7].

There are a lot of methods for surface waves excitation: they can be excited either by near field (e.g. from radiating molecules or quantum dots near the surface along which the surface wave propagates [8] or from a probe of a near-field optical microscope [9]) or can be excited by far field (of a laser beam incident on the surface). In the last case, effective surface-wave excitation requires the equality of the surface wave's and the incident exciting wave's wave-vector components along the surface (which is achieved in the Kretschmann configuration). Alternatively, structures breaking translational symmetry along the propagation direction of the surface wave should be fabricated. For instance, the translational symmetry may be broken by a periodic grating on the surface along which the surface wave propagates.

Surface waves are commonly excited by means of periodic gratings [10]. The efficiency of a surface-wave excitation on a periodic grating may be, however, not high enough particularly in the case of a small grating with just several periods. To increase the efficiency of a surface-wave excitation while keeping an exciting structure (e.g. grating) small (for the purpose of miniaturization) new approaches to develop such structures should be found.

In our paper [11], by the example of a surface plasmonic wave (PW) in a dielectric layer on metal we show that holographic principles can be used for efficient excitation of the PWs. In some particular cases, the efficiency of holograms may be substantially higher [11] than that of simple periodic gratings in the dielectric layer (e.g. similar to those described in [10]) in which the refractive index is periodically modulated in the layer plane and is invariable in the direction perpendicular to the layer plane. In the present paper, we expand our study of holographic excitation of optical surface waves.

Let us illustrate the holographic surface-waves excitation method by the example of a surface plasmon-polariton (SPP) excitation on a sinusoidal metal surface relief modulation. In this case, the grating in the form of a metal surface relief modulation may be considered as an analog of a hologram. The SPP being excited is an analog of either an object wave used for a hologram recording or an image obtained as a result of a hologram reconstruction; an exciting wave incident on the grating from air is an analog of either a reference or a reconstructing wave. The configuration for SPPs excitation using principles of holography is shown in Fig. 1.

Let us consider, according to Fig. 1, excitation of a SPP $\mathbf{E}_{\text{SPP}}(y)\exp(ik_{\text{SPP}}x)$ by a plane wave $\mathbf{E}_{\text{inc}}(y)\exp(ik_{\parallel}x)$ on a hologram in the form of a metal surface relief modulation (here $k_{\parallel}$ is the *x*-component of the incident plane-wave wave vector). The exciting *p*-polarized plane wave is incident in the *xy*-plane. Suppose the metal surface relief in the hologram area (of length $L$) is modulated such that the *y*-coordinate of the surface for each *x* is determined by the cross interference term of the exciting (reference) wave $\mathbf{E}_{\text{exc}}(y)\exp(ik_{\parallel}x)$ (being the sum of the incident wave $\mathbf{E}_{\text{inc}}(y)\exp(ik_{\parallel}x)$ and the corresponding reflected wave without a hologram) and the SPP, used for the hologram recording. That is, in the hologram area the metal surface relief is described by a function

$$y(x) \propto \text{Re}\left\{\left(\mathbf{E}_{\text{exc}}(0)\cdot\mathbf{E}^{*}_{\text{SPP}}(0)\right)\exp\left[i\left(k_{\parallel}-k_{\text{SPP}}\right)x\right]\right\}$$

$$\propto \cos\left[(k_{\parallel}-k_{\text{SPP}})x+\varphi\right].$$

Here $\varphi$ is a phase shift (independent on *x*) between the exciting wave and the SPP. Thus, the hologram obtained in the form of the metal surface relief modulation $y(x)$ is the grating with the period $D$, given by the condition (so called Wood's condition [12-14])

$$\mathbf{k}_\parallel + m\mathbf{G} \approx \text{Re}\,\mathbf{k}_{\text{SPP}}. \qquad (1)$$

Here $\mathbf{G}$ is the reciprocal lattice vector ($D = 2\pi/G$), $m$ is an integer (in the case of the hologram considered, all the vectors in (1) are parallel to the $x$-axis and $m = 1$). As is known [12,13], a $p$-polarized plane wave $\mathbf{E}_{\text{inc}}(y)\exp(ik_\parallel x)$ (coincident with that used for the grating recording) incident on such a grating excites the SPP $\mathbf{E}_{\text{SPP}}(y)\exp(ik_{\text{SPP}}x)$ (Wood effect). Thus, in the simple case of an SPP excitation on a periodic metal surface modulation such a periodic modulation may be considered as a hologram. However, in the more general case of excitation of optical surface waves with sophisticated field distributions or in the case of exciting waves not being plane the described holographic approach can be applied as well.

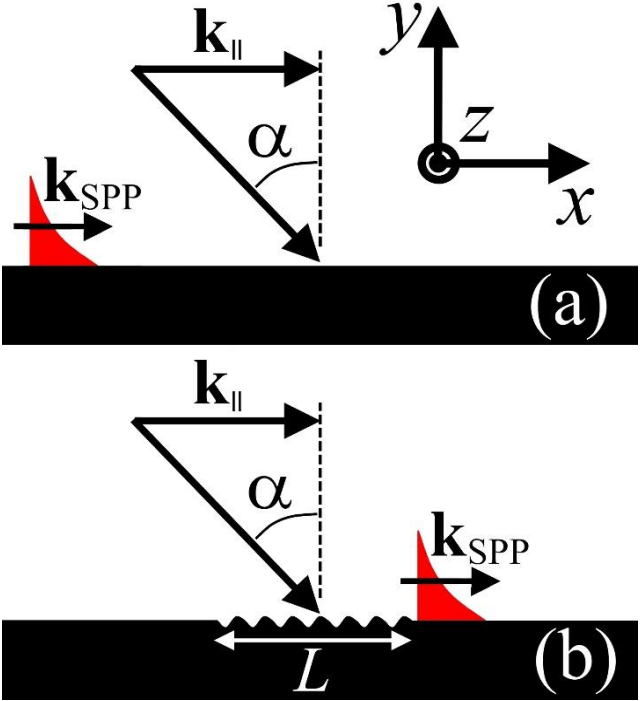

**Fig. 1.** Hologram recording (Fig. a) on a metal surface and hologram reconstruction (Fig. b) for an SPP excitation by far field. (a) For the hologram recording on the metal surface one must create an interference pattern of a reference wave incident from air at the angle $\alpha$ and the SPP (with the wave vector $\mathbf{k}_{\text{SPP}}$). (b) The hologram in the form of modulated metal surface is illuminated by a reconstructing wave (exactly matching the reference wave). As a result of the reconstructing wave scattering on the hologram, the SPP is excited. Air is shown by white color, metal substrate – by black color.

In the recent years, a holographic approach for excitation of various complex-configuration plasmonic beams by laser beams incident from air [15-18] or by simple SPPs [19-21] is being actively developed. Holograms for excitation of plasmonic Airy [15,16,18,22], plasmonic Hermite-Gauss [15,18,23], plasmonic Bessel [18] beams on a flat metal surface and other plasmonic beams with rather complex field distributions [20] are developed and fabricated. One class of plasmonic holograms proposed in literature includes holograms in a form of metal surface modulation determined by interference of an exciting wave with a plasmonic beam being excited [15,18,24-26]. Holograms of another class are metasurfaces composed of optical nanoantennas whose characteristics and arrangement are customized to create a required amplitude-phase distribution of the scattered field [16,17,20,23]. However, application of holography for efficient (in terms of power) excitation of plasmonic surface waves in metal-dielectric structures and modes of plasmonic waveguides is still to be analyzed.

In the present paper, we will theoretically study efficiency of surface-waves excitation using principles of holography. By the particular example of a plasmonic wave (PW) excitation in the structure metal/dielectric layer/air using volume holograms in the dielectric layer, we will find out general enough conditions when the holographic approach can be efficient for surface waves excitation. We will discuss advantages of the holograms over simple periodic gratings where material parameters are modulated in the plane of a substrate and are invariable in the direction perpendicular to the substrate plane. In addition, we will consider a possibility to increase efficiency of holograms by recording of them in several successive iterations.

## 2. GEOMETRY OF THE SYSTEM UNDER CONSIDERATION AND HOLOGRAMS FOR EXCITATION OF PLASMONIC WAVES IN THE SYSTEM

In this paper, we analyze application of holography principles for a PW excitation in a dielectric layer on metal. In particular, the structure analyzed here was composed of a layer (of infinite width) of photoresist poly(methyl methacrylate) (PMMA) with the thickness $h = 600$ nm on a gold surface (see Fig. 2(a)). All the quantitative results presented in this paper are for the wavelength in vacuum $\lambda = 1.55$ μm. For this wavelength the PMMA refractive index was taken to be $n_{\text{PMMA}} = 1.481$ [27] and the dielectric permittivity of gold – to be $\varepsilon_{\text{Au}} = -115.1 + 11.3i$ [28]. In the PMMA layer on gold at $\lambda = 1.55$ μm there is only one TM-polarized guided mode which is the PW that is considered in the present paper. The wave number of the PW is $k_{\text{PW}} = (1.433 + 0.003i)k_0$. All the numeric results in the present paper were obtained with Comsol Multiphysics.

As an exciting wave, we considered a beam $\mathbf{E}_{\text{inc}}\exp(ik_\parallel x)$ incident in the $xy$-plane (see the coordinate system in Fig. 2(a)). The beam had the Gaussian profile in the plane of incidence (having the focus on the gold surface) and $\mathbf{E}_{\text{inc}}$ independent on $z$; $\mathbf{E}_{\text{inc}}$ was polarized in the plane of incidence. The beam waist (in the $xy$-plane) was $w_0 = 6\lambda$. $k_\parallel = k_0 \sin\alpha$, where $k_0 = 2\pi/\lambda$ and $\alpha$ is the incidence angle of the exciting beam.

A hologram was recorded in the rectangular area of the PMMA layer (we denote this area as $\Omega$) which covered the whole thickness of the layer in the $y$-direction, was infinite in the $z$-direction and had the length $L = 7.0$ μm in the $x$-direction (see Fig. 2(a,b)). The center of the area $\Omega$ along the $x$-direction was at the incident beam's focus.

For holograms recording, we considered an interference of the exciting beam $\mathbf{E}_{\text{exc}}\exp(ik_\parallel x)$ (which is the sum of the incident beam $\mathbf{E}_{\text{inc}}\exp(ik_\parallel x)$ and the corresponding reflected beam before the recording) and the plasmonic wave $\mathbf{E}_{\text{PW}}\exp(ik_{\text{PW}}x)$ propagating in the $x$-direction. The shape of holograms (recorded in the area $\Omega$) was determined by the cross interference term $\text{Re}\left\{\left(\mathbf{E}_{\text{exc}}\cdot\mathbf{E}_{\text{PW}}^*\right)\exp\left[i\left(k_\parallel - k_{\text{PW}}\right)x\right]\right\}$ in the intensity of the exciting beam and the PW sum. We considered the holograms with the continuously varying refractive index $n(x,y)$ in the area $\Omega$:

$$n(x,y) = n_{PMMA} + A(n_{PMMA} - 1) f(x,y)/\max_{\Omega}(|f(x,y)|) \quad \text{if } A|f(x,y)|/\max_{\Omega}(|f(x,y)|) \leq 1,$$
$$n(x,y) = n_{PMMA} + (n_{PMMA} - 1) = 2n_{PMMA} - 1 \quad \text{if } A f(x,y)/\max_{\Omega}(|f(x,y)|) > 1, \quad (2)$$
$$n(x,y) = n_{PMMA} - (n_{PMMA} - 1) = 1 \quad \text{if } A f(x,y)/\max_{\Omega}(|f(x,y)|) < -1.$$

Here $f(x,y)$ denotes $\text{Re}\{(\mathbf{E}_{exc} \cdot \mathbf{E}_{PW}^*) \exp[i(k_\parallel - k_{PW})x]\}$. The maxima were found over the whole area $\Omega$ of a hologram, the parameter $A$ described a hologram "contrast degree" and ranged from 0 to $+\infty$. Below, we optimize holograms with respect to the $A$ parameter in order to maximize the efficiency of the PW excitation. Thus, as is clear from the definition (2), the refractive index $n(x,y)$ in $\Omega$ could have values within the range from 1 to $2n_{PMMA} - 1$. For $A \leq 1$ the deviation of the refractive index from its initial value $n_{PMMA}$ before a hologram recording not exceeded $A(n_{PMMA} - 1)$. In the limit $A \to +\infty$ a hologram became discrete with $n(x,y) = 2n_{PMMA} - 1$ at $f(x,y) > 0$ and $n(x,y) = 1$ at $f(x,y) < 0$.

The hologram obtained in this way for the exciting wave incident at the angle of $\alpha = -45°$ (that is with $k_\parallel = k_0 \sin\alpha < 0$) and with $A = 2.0$ is shown in Fig. 2(b). A refractive index distribution in the PMMA layer after recording of holograms in $\Omega$ is shown by color in the figure. Everywhere below we consider only $k_\parallel \leq 0$ (i.e. $\alpha \leq 0$) because for negative $\alpha$ (and positive $\text{Re}\,k_{PW}$) only one side diffraction lobe scattered on a hologram coupled to the excited PW while the rest ("parasitic") were nonpropagating. Indeed, from the condition of the type (1) for the diffraction lobes with $m \leq -1$ $|k_\parallel + mG| \approx \text{Re}\,k_{PW} + 2|k_\parallel| + |m+1|G > \text{Re}\,k_{PW}$ and for $m \geq 2$ $|k_\parallel + mG| \approx \text{Re}\,k_{PW} + (m-1)G > \text{Re}\,k_{PW}$.

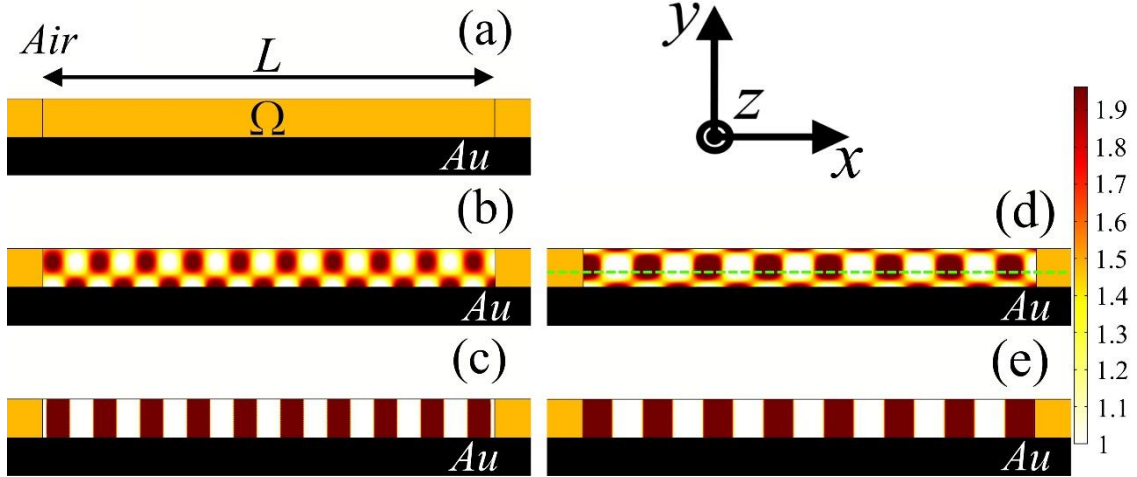

**Fig. 2.** (a) The structure with a PMMA layer on gold. The black area corresponds to a gold substrate, the gray area (yellow online) – to a PMMA layer, the white area is air with the refractive index 1. In the area $\Omega$ of length $L$ inside the PMMA layer a hologram/grating will be recorded. (b) In the area $\Omega$ the hologram of the form (2) for the exciting wave incident at the angle $\alpha = -45°$ with $A = 2.0$ is recorded. (c) In the area $\Omega$ the grating of the form (3) with the period $D = 724$ nm and with $A = 10.0$ is created. (d) In the area $\Omega$ the hologram for the exciting wave incident at the angle $\alpha = -13°$ with $A = 3.0$ is recorded. The horizontal green dashed line indicates $y = 234$ nm at which $E_{excy}$ changes its sign. (e) In the area $\Omega$ the grating with the period $D = 935$ nm and with $A = 10.0$ is created. The color scale to the right indicating refractive index in the PMMA layer is common for Figs. 2(a)-2(e).

For comparison with holograms efficiency, we studied excitation of the PW on simple periodic gratings in the PMMA layer with refractive index modulated in the layer plane and invariable in the direction perpendicular to the layer plane. The gratings were created in the area $\Omega$ exactly matching (by shape and position) that one where holograms were recorded (Fig. 2(a)). The geometries of the gratings considered are shown in Figs. 2(c,e) where the refractive index is indicated by the same colors as in Fig. 2(b). In general, the grating period $D$ optimal with respect to the PW excitation efficiency may differ from that given by the condition of the type (1) because the condition of the type (1) is valid in the low-contrast limit (when the refractive index in $\Omega$ is weakly modulated and the mean refractive index is close to $n_{PMMA}$). However, for the gratings considered in the current paper, the mean refractive index in $\Omega$ was nearly equal to $n_{PMMA}$ resulting in most cases in the optimal grating period being close to that given by the condition of the type (1). So, we considered gratings in the area $\Omega$ with the refractive index $n(x)$ independent on $y$:

$$n(x) = n_{PMMA} - A(n_{PMMA}-1)\cos(Gx) \quad \text{if } A|\cos(Gx)| \leq 1,$$
$$n(x) = n_{PMMA} + (n_{PMMA}-1) = 2n_{PMMA} - 1 \quad \text{if } A\cos(Gx) < -1, \quad (3)$$
$$n(x) = n_{PMMA} - (n_{PMMA}-1) = 1 \quad \text{if } A\cos(Gx) > 1.$$

As in the case of holograms, here the parameter $A$ described a gratings "contrast degree" and took values from 0 to $+\infty$. We optimized the gratings with respect to both the $A$ parameter and the period $D = 2\pi/G$ in order to maximize the efficiency of the PW excitation. The refractive index $n(x)$ in $\Omega$ could have values within the range from 1 to $2n_{PMMA}-1$. For $A \leq 1$ the deviation of the refractive index from $n_{PMMA}$ not exceeded $A(n_{PMMA}-1)$. In the limit $A \to +\infty$ a grating became discrete with the refractive indices of $2n_{PMMA}-1$ and 1.

## 3. EFFICIENCY OF PLASMONIC WAVE EXCITATION USING THE VOLUME HOLOGRAMS

We compared efficiencies $\eta$ of the proposed holograms (2) with those of the gratings (3) at various incidence angles of an exciting wave in the range from $\alpha = 0°$ to $\alpha = -70°$. The holograms and gratings efficiency $\eta$ was defined as a ratio of the excited PW power to the right of the grating/hologram to the total exciting wave power. The calculated efficiencies of the gratings and the holograms as functions of $\alpha$ are shown in Fig. 3. We optimized the efficiencies $\eta$ of the gratings/holograms for each $\alpha$ (the holograms were optimized with respect to $A$, the gratings – with respect to $A$ and the period $D$).

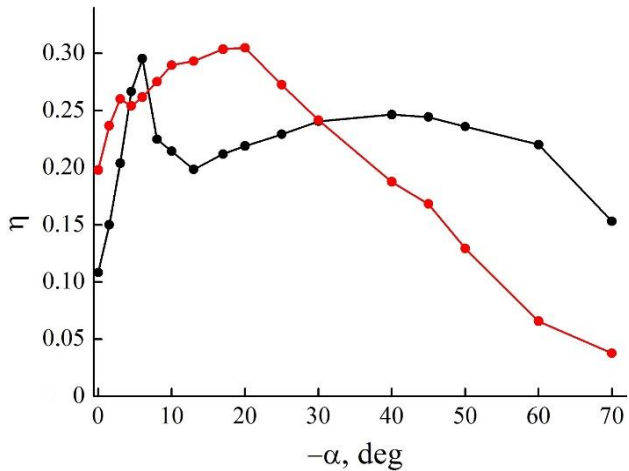

**Fig. 3.** Efficiencies $\eta$ of the plasmonic mode (PW) excitation in the PMMA layer on gold by the use of holograms (2) (black line) and periodic gratings (3) (gray (red online) line) as functions of the incidence angle $\alpha$. The holograms/gratings were optimized for each $\alpha$. The calculated efficiencies are indicated by circles, between the circles straight lines are drawn.

As follows from Fig. 3, the holograms (2) are considerably more efficient than the gratings (3) for oblique incidence of the exciting wave ($|\alpha| > 30°$). However, at the near-zero angles $\alpha$ and for $-30° < \alpha < -7°$ the gratings are more efficient. Below we analyze the cases of $\alpha = -45°$, $\alpha = 0°$ and $\alpha = -13°$ in more details in order to see special features of holograms for surface-waves excitation and find out general conditions when holograms can be more efficient than simple periodic gratings (where material parameters are invariable in the direction perpendicular to the substrate plane).

First, let us consider the case of $\alpha = -45°$. In this case, the efficiency of the optimized (with $A = 2.0$) hologram is 24.4% and the efficiency of the optimized grating (with $A = 10.0$ and $D = 724$ nm) is 16.8%. In Figs. 4(a,b), the electric field distributions for the optimized hologram and grating illuminated by an incident beam are shown. The excited PW is propagating to the right of the hologram/grating. In Fig. 2(b), one can see that the refractive index distribution in the upper part (at $y > 267$ nm) of the hologram is shifted along the $x$-direction by a half-period relative to the lower part. Obviously, such a shift is absent in the simple grating (3). Let us show qualitatively that it is this shift that makes the hologram more efficient than the grating. Consider the field distribution of the exciting wave and the PW in the area $\Omega$. Numerical calculation showed that the components $E_{PWx}\exp(ik_{PW}x)$ and $E_{PWy}\exp(ik_{PW}x)$ in $\Omega$ had phase fronts nearly flat (perpendicular to the $x$-axis). On the other hand, the phase fronts shape of $\mathbf{E}_{exc}\exp(ik_\parallel x)$ is more complex: due to large $|\text{Re}\,\varepsilon_{Au}|$ the dependence of $E_{excy}$ on the $y$-coordinate may be approximately described by the function $\cos\{[(k_0 n_{PMMA})^2 - k_\parallel^2]^{1/2} y\}$ having a maximum at $y = 0$ (on the gold surface) and changing its sign at $y = 296$ nm (in reality $E_{excy}$ changes sign at $y \approx 267$ nm). Thus, for the single scattering of the component $E_{excy}\exp(ik_\parallel x)$, the $y$-polarized radiation scattered from the lower and the upper parts of the hologram in the $x$-direction is in phase with each other. Therefore, the PW is efficiently excited on the hologram. The dependence of the component $E_{excx}$ on the $y$-coordinate may be approximately described by the function $\sin\{[(k_0 n_{PMMA})^2 - k_\parallel^2]^{1/2} y\}$, that is, phase fronts of the component $E_{excx}\exp(ik_\parallel x)$ are flat almost everywhere in $\Omega$. However, scattering of the component $E_{excx}\exp(ik_\parallel x)$ on the hologram has seemingly low impact on the PW excitation because $x$-polarized induced dipoles radiate weakly in the $x$-direction and because $x$- and $y$-components of $\mathbf{E}_{exc}$ are comparable $(|E_{excx}|/|E_{excy}| = (n_{PMMA}^2 - \sin^2\alpha)^{1/2}/\sin\alpha = 1.84)$.

In the case of $E_{excy}\exp(ik_\parallel x)$ scattering on the grating (3), the radiation scattered in the $x$-direction on the grating lower part (at $y < 267$ nm) is out of phase with that scattered on the grating upper part.

Thus, based on the case $\alpha = -45°$ considered, we can conclude that a hologram would be more efficient than a grating of the form (3) if the refractive index distribution in the hologram is considerably inhomogeneous in the $y$-direction. In particular, the $n(x,y)$ can be considerably dependent on $y$ in $\Omega$ if the exciting wave is incident obliquely at a large angle α or when the exciting wave has a more complex field distribution in $\Omega$.

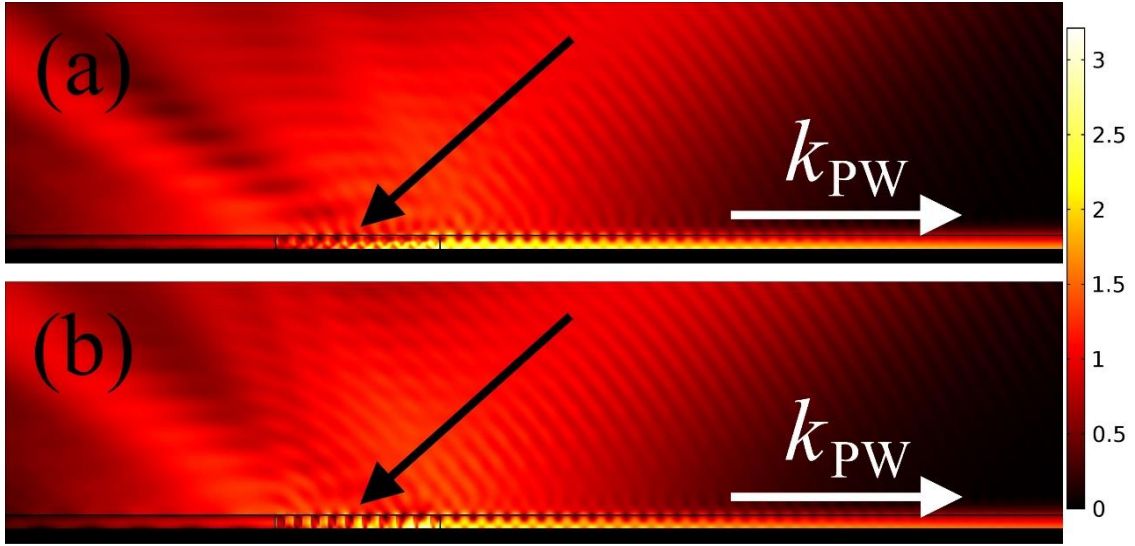

**Fig. 4.** (a) The electric field magnitude distribution for the optimized hologram (2) with $A = 2.0$ illuminated by the exciting beam at $\alpha = -45°$ (the propagation direction of the incident beam is indicated by the black diagonal arrow). The area of the hologram is situated under the black arrow tip. The excited plasmonic mode (PW) of the PMMA layer on gold is propagating to the right of the hologram. (b) The electric field magnitude distribution for the optimized grating of the form (3) (with $A = 10.0$ and $D = 724$ nm) illuminated by the exciting beam at $\alpha = -45°$. The color scale to the right (in arbitrary units) is common for Figs. 4(a) and 4(b).

Now consider the case of a normal incidence of the exciting wave. In this case the efficiency of the optimized (with $A = 1.3$) hologram is 10.8% and the efficiency of the optimized grating (with $A = 1.3$ and $D = 1027$ nm) is 19.8%. The refractive index distribution $n(x, y)$ in the hologram determined by the value
$\mathrm{Re}\left\{(\mathbf{E}_{\mathrm{exc}} \cdot \mathbf{E}_{\mathrm{PW}}^*) \exp\left[i(k_\| - k_{\mathrm{PW}})x\right]\right\}$
$\approx \mathrm{Re}\left[ E_{\mathrm{exc}x} E_{\mathrm{PW}x}^* \exp(-i k_{\mathrm{PW}} x) \right]$
$\propto \sin(k_0 n_{\mathrm{PMMA}} y) \cos(k_{\mathrm{PW}} x + \varphi)$ is similar to that in the grating in the sense that the sign of $n(x, y) - n_{\mathrm{PMMA}}$ is independent on $y$ almost everywhere in $\Omega$. However, the optimal grating period $D = 1027$ nm is different from the hologram "period" given by the condition of the type (1) and equal to $D = 1082$ nm. Apparently, the difference of the optimal grating period from $D = 1082$ nm along with the independence of the $n(x, y) - n_{\mathrm{PMMA}}$ sign on $y$ (as well as in the grating) almost everywhere in the hologram results in the advantage of the grating over the hologram. Really, if we take a grating of the type (3) with $D = 1082$ nm and optimize it with respect to only the parameter $A$ the efficiency of the optimized grating would be only 10.0% (less than the efficiency of the optimized hologram).

From the case of normal incidence considered, we can infer that holograms may be less efficient than gratings of the type (3) if the optimal grating period differs from the period given by the relation of the type (1) (whereas a hologram "period" in the x-direction is always close to that given by the relation of the type (1)). Nevertheless, as we show in [11], a hologram can be more efficient than a grating with the optimal period significantly different from the hologram "period" when the exciting wave is incident obliquely at a rather large angle.

Finally, we analyze the case of $\alpha = -13°$. In this case, the optimal grating period $D = 935$ nm and the period given by the relation of the type (1) are the same. However, the optimized grating is more efficient than the optimal hologram. In order to discover why the hologram is less efficient than the grating, we examine the refractive index distribution in the hologram (see Fig. 2(d)) and the distribution of $\mathbf{E}_{\mathrm{exc}}$ in $\Omega$. At $\alpha = -13°$ the contribution of the term $\mathrm{Re}\left\{E_{\mathrm{exc}x} E_{\mathrm{PW}x}^* \exp[i(k_\| - k_{\mathrm{PW}})x]\right\}$ into the function $n(x, y)$ is comparable with that of $\mathrm{Re}\left\{E_{\mathrm{exc}y} E_{\mathrm{PW}y}^* \exp[i(k_\| - k_{\mathrm{PW}})x]\right\}$ (because $\max_\Omega[\mathrm{Re}(E_{\mathrm{exc}x} E_{\mathrm{PW}x}^*)] / \max_\Omega[\mathrm{Re}(E_{\mathrm{exc}y} E_{\mathrm{PW}y}^*)] = 1.5$). This leads to the change of the $n(x, y) - n_{\mathrm{PMMA}}$ sign at $y \approx 89$ nm considerably different from $y \approx 234$ nm (marked with green dashed line in Fig. 2(d)) where $E_{\mathrm{exc}y}$ changes its sign. Therefore, for the single scattering of the component $E_{\mathrm{exc}y} \exp(i k_\| x)$, the radiation scattered in the x-direction from the hologram parts $y < 89$ nm and $234\ \mathrm{nm} < y < 500$ nm is out of phase with the radiation scattered from the parts $89\ \mathrm{nm} < y < 234$ nm and $y > 500$ nm. By analogy, for the scattering of $E_{\mathrm{exc}x} \exp(i k_\| x)$ in the x-direction, the radiation scattered from the hologram part $y < 89$ nm is out of phase with the radiation scattered from the part $y > 89$ nm (because $E_{\mathrm{exc}x}$ changes its sign only at $y \approx 500$ nm).

Seemingly, the specified disagreement between the refractive index distribution in the hologram and the phase fronts shape of $E_{\mathrm{exc}x} \exp(i k_\| x)$ and $E_{\mathrm{exc}y} \exp(i k_\| x)$ results in the relatively low efficiency of holograms at $-30° < \alpha < -7°$. To confirm this supposition, we modified holograms by taking the value

$$\mathrm{Re}\left\{\left(\frac{E_{\mathrm{exc}x}E_{\mathrm{PW}x}^*}{|E_{\mathrm{PW}x}|}+\frac{E_{\mathrm{exc}y}E_{\mathrm{PW}y}^*}{|E_{\mathrm{PW}y}|}\right)\exp[i(k_\parallel-k_{\mathrm{PW}})x]\right\}$$

as a function $f(x,y)$ in (2). In such a modified hologram, at $\alpha=-13°$ the summand $\mathrm{Re}\{E_{\mathrm{exc}x}E_{\mathrm{PW}x}^*\exp[i(k_\parallel-k_{\mathrm{PW}})x]/|E_{\mathrm{PW}x}|\}$ is dominant ($|E_{\mathrm{exc}x}|/|E_{\mathrm{exc}y}|=6.5$). In Fig. 5, efficiencies of the modified holograms optimized at each $\alpha$ are shown by the green line. For the angles $-14°<\alpha<-6°$, the efficiencies of the modified holograms are considerably higher than the efficiencies of the original-type holograms and exceed the gratings (3) efficiencies as well.

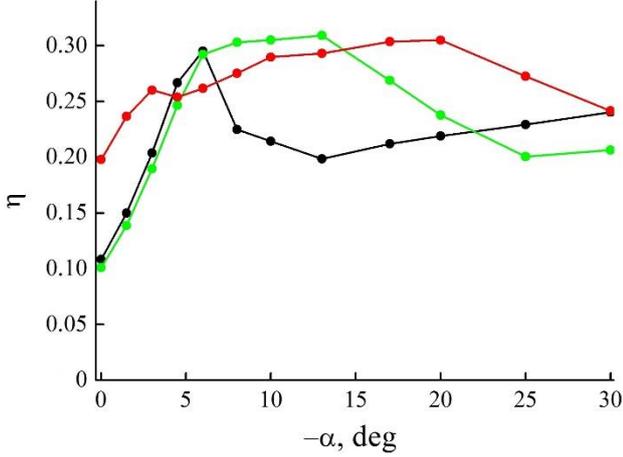

**Fig. 5.** Efficiencies $\eta$ of the plasmonic mode (PW) excitation in the PMMA layer on gold by the use of holograms (2) (black line), periodic gratings (3) (dark-gray (red online) line) and modified holograms (light-gray (green online) line) as functions of the incidence angle $\alpha$. The holograms/gratings were optimized for each $\alpha$. The calculated efficiencies are indicated by circles, between the circles straight lines are drawn. Black and dark-gray (red) lines are the same as in Fig. 3.

Thus, in the case when the terms $\mathrm{Re}\{E_{\mathrm{exc}x}E_{\mathrm{PW}x}^*\exp[i(k_\parallel-k_{\mathrm{PW}})x]\}$ and $\mathrm{Re}\{E_{\mathrm{exc}y}E_{\mathrm{PW}y}^*\exp[i(k_\parallel-k_{\mathrm{PW}})x]\}$ make comparable contributions into the function $n(x,y)$ and spatial distributions of $\mathrm{Re}\{E_{\mathrm{exc}x}E_{\mathrm{PW}x}^*\exp[i(k_\parallel-k_{\mathrm{PW}})x]\}$ and $\mathrm{Re}\{E_{\mathrm{exc}y}E_{\mathrm{PW}y}^*\exp[i(k_\parallel-k_{\mathrm{PW}})x]\}$ are different in $\Omega$ the efficiency of a hologram (2) may decrease and become lower than that of gratings.

## 4. HOLOGRAMS RECORDED IN SEVERAL ITERATIONS

Strictly speaking, holography is a method working in the single-scattering approximation. Otherwise, the radiation multiply scattered on a hologram may in general be completely different from an object wave which determined the function $f(x,y)$ and the shape of the hologram. In particular, as was shown in the previous section for the $\alpha=0°$ case and the large-contrast holograms/gratings, the optimal grating period may considerably differ from the hologram "period" (given by the condition of the type (1) and optimal in the low-contrast limit). Below we analyze if it is possible to increase efficiency of holograms by recording of them in several successive iterations such that only small refractive index variations are made at each iteration. At each new iteration the hologram is recorded in the same area $\Omega$. The refractive index variation at a current iteration is determined by the interference of the PW being excited and an exciting wave. In turn, the exciting wave changes from iteration to iteration, since it is the result of the incident wave scattering on the hologram recorded at the previous iteration.

We considered holograms recorded in two successive iterations in more details. After the first iteration, the refractive index distribution $n(x,y)$ in $\Omega$ had the form (2) with some value of the $A$ parameter and with $f(x,y)=\mathrm{Re}\{(\mathbf{E}_{\mathrm{exc}}\cdot\mathbf{E}_{\mathrm{PW}}^*)\exp[i(k_\parallel-k_{\mathrm{PW}})x]\}$. As before, $\mathbf{E}_{\mathrm{exc}}\exp(ik_\parallel x)$ is the sum of the incident Gaussian beam and the corresponding reflected wave in the case of homogeneous PMMA layer without a hologram. Again, $\mathbf{E}_{\mathrm{PW}}\exp(ik_{\mathrm{PW}}x)$ is the plasmonic wave being excited. After the second iteration of recording, the refractive index distribution $n_2(x,y)$ in $\Omega$ was

$$
\begin{aligned}
n_2(x,y) &= n(x,y)+A_2\left(n_{\mathrm{PMMA}}-1\right)\frac{f_2(x,y)}{\max_\Omega(|f_2(x,y)|)} && \text{if } \left|n(x,y)+A_2\left(n_{\mathrm{PMMA}}-1\right)\frac{f_2(x,y)}{\max_\Omega(|f_2(x,y)|)}-n_{\mathrm{PMMA}}\right|\le n_{\mathrm{PMMA}}-1,\\
n_2(x,y) &= 2n_{\mathrm{PMMA}}-1 && \text{if } n(x,y)+A_2\left(n_{\mathrm{PMMA}}-1\right)\frac{f_2(x,y)}{\max_\Omega(|f_2(x,y)|)}-n_{\mathrm{PMMA}}>n_{\mathrm{PMMA}}-1, \qquad (4)\\
n_2(x,y) &= 1 && \text{if } n(x,y)+A_2\left(n_{\mathrm{PMMA}}-1\right)\frac{f_2(x,y)}{\max_\Omega(|f_2(x,y)|)}-n_{\mathrm{PMMA}}<-(n_{\mathrm{PMMA}}-1).
\end{aligned}
$$

Here $n(x,y)$ is the refractive index distribution after the first iteration, $A_2$ is a constant (possessing values in the range from 0 to $+\infty$), $f_2(x,y)=\mathrm{Re}[(\mathbf{E}_{\mathrm{exc2}}\cdot\mathbf{E}_{\mathrm{PW}}^*)\exp(-ik_{\mathrm{PW}}x)]$. $\mathbf{E}_{\mathrm{exc2}}$ is the sum of the incident Gaussian beam and the wave scattered from the PMMA layer on gold with the hologram (2) recorded at the first iteration. As in the case of holograms (2), the refractive index $n_2(x,y)$ in the two-iteration holograms (4) could have values within the range from 1 to $2n_{\mathrm{PMMA}}-1$.

The two-iteration hologram for the normally incident exciting Gaussian beam is shown in Fig. 6 (for $A=0.6$ and $A_2=7.0$). The color scale is the same as in Fig. 2.

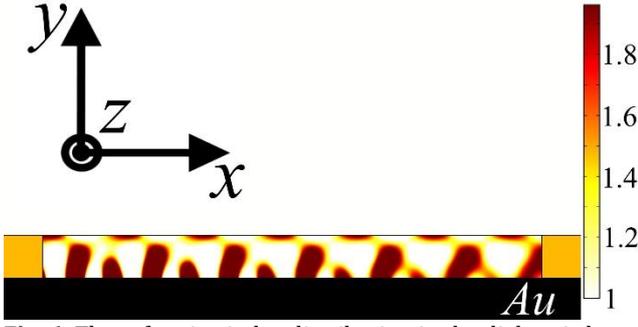

**Fig. 6.** The refractive index distribution in the dielectric layer with the two-iteration hologram (4) with $A = 0.6$ and $A_2 = 7.0$. The exciting beam is incident normally to the layer plane. The black area corresponds to the gold substrate. The color scale to the right indicates the refractive index values in the dielectric layer.

Note that at the first and the second iterations of recording the phase relations between the exciting wave and the wave being excited must be the same. If $\mathbf{E}_{\text{PW}}$ entering the expression for $f(x, y)$ is the same (in particular, having the same phase) as $\mathbf{E}_{\text{PW}}$ entering the expression for $f_2(x, y)$, the phase of the incident Gaussian beam in $\mathbf{E}_{\text{exc}}\exp(ik_\parallel x)$ must be equal to that of the incident Gaussian beam in $\mathbf{E}_{\text{exc2}}$. This requirement can be easily understood in the limit of small refractive index variation at each iteration (in the limit of small $A$ and $A_2$). In this limit due to small scattering of the incident wave on the first-iteration hologram, $\mathbf{E}_{\text{exc2}}$ is almost equal to $\mathbf{E}_{\text{exc}}\exp(ik_\parallel x)$. If the phase relations between the excited wave and the PW are the same for the first and the second iterations, $f_2(x, y) \approx f(x, y)$ and the refractive index variation $A(n_{\text{PMMA}} - 1) f(x, y) / \max_\Omega (|f(x, y)|)$ at the first iteration is equal (within a constant positive factor) to the variation $A_2 (n_{\text{PMMA}} - 1) f_2(x, y) / \max_\Omega (|f_2(x, y)|)$ at the second iteration. As a result, the PW excited due to scattering on $A(n_{\text{PMMA}} - 1) f(x, y) / \max_\Omega (|f(x, y)|)$ interferes constructively with the PW excited due to scattering on $A_2 (n_{\text{PMMA}} - 1) f_2(x, y) / \max_\Omega (|f_2(x, y)|)$.

## 5. PLASMONIC WAVES EXCITATION EFFICIENCY OF TWO-ITERATIONS RECORDED HOLOGRAMS

We calculated efficiencies η of the two-iteration holograms (4) for various incidence angles α. These two-iteration holograms were optimized by selecting the values of $A$ and $A_2$ in order to maximize their efficiencies. The obtained dependence is shown in Fig. 7.

As can be seen in Fig. 7, the two-iteration holograms (4) are substantially more efficient than the optimized single-iteration holograms (2) and the optimized gratings (3) for all α considered.

Note the particular case of $\alpha = 0°$. The refractive index distribution of the optimized two-iteration hologram (with $A = 0.6$ and $A_2 = 7.0$) is shown in Fig. 6. As opposed to single-iteration holograms (2) and gratings (3), the refractive index distribution in Fig. 6 is highly asymmetric with respect to the inversion of the *x*-axis direction. The power of the PW excited in the *x*-direction on the two-iteration hologram is 15 times more than the power of a PW excited in the opposite direction whereas in the case of the hologram (2) and the grating (3) the power of the excited PWs is nearly equal in both directions. The efficiency of the two-iteration hologram is 2.9 times more than the efficiency of the single-iteration hologram (2) at $\alpha = 0°$.

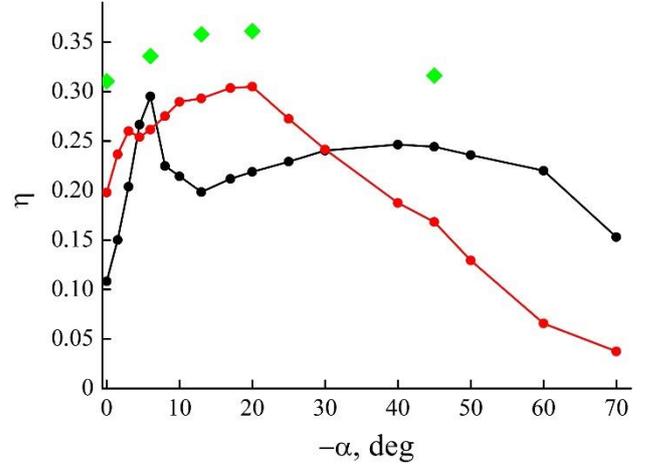

**Fig. 7.** Efficiency η of the PW excitation in the PMMA layer on gold by the use of continuous two-iteration holograms (4) (light-gray (green online) diamonds) as a function of the incidence angle α. The holograms were optimized (by selection of $A$ and $A_2$) for each α. The black line with circles and the dark-gray (red online) line with circles correspond to the holograms (2) and the gratings (3) (these lines are the same as in Fig. 3 and are shown for comparison).

## 6. CONCLUSION

In the present paper, we theoretically studied capabilities of the holography principle in application for efficient excitation of plasmonic surface waves by an incident laser beam.

It is easy to see that commonly known periodic gratings in the form of a metal surface relief modulation used for excitation of SPPs on a metal surface can be considered as holograms. However, the holographic approach can potentially be generalized for excitation of various optical surface waves and optical waveguide modes. In particular, for the telecommunication wavelength, we analyzed holograms for excitation of a plasmonic surface wave in a structure gold/subwavelength PMMA layer/air. To be more specific, volume holograms in the form of the refractive index modulation in the PMMA layer were considered. The refractive index distribution in these holograms might be significantly inhomogeneous in the *y*-direction perpendicular to the layer plane, because of complicated (curved) phase fronts shape of an exciting wave in the hologram area.

We studied efficiencies (as a ratio of the excited plasmonic wave power to the exciting incident beam total power) of the proposed holograms as functions of the exciting Gaussian beam incidence angle. For comparison, we considered efficiencies of simple periodic gratings (of the type (3)) with refractive index of the PMMA layer modulated in the layer plane and invariable in the *y*-direction. The gratings/holograms considered were optimized for each incidence angle individually. The results obtained show that at some incidence angles the holograms (2) are more efficient than the gratings (especially at large incidence angles) and at some other incidence angles the gratings are more efficient. For the incidence angle –45° the optimized hologram efficiency is 1.5 times higher than the optimized grating efficiency. For more oblique incidence angles the advantage of holograms over gratings is even more sizeable.

Based on the numerical analysis, we found out factors determining advantages and disadvantages of holograms (2) with respect to gratings (3). Holograms "periods" were determined by the condition of the type (1) and might differ from the optimal gratings periods if the refraction index modulation was not weak (strictly speaking, holograms are intended to work in the single-scattering regime). However, even if an optimal grating period considerably differs from a hologram "period", the hologram can be more efficient if the refractive index distribution in the hologram is significantly inhomogeneous in the $y$-direction [11]. For example, the refractive index distribution in a hologram is considerably different from that in simple gratings (independent on $y$) if the exciting wave is incident on a hologram obliquely at a rather large angle or if phase fronts of either exciting wave or a wave being excited are curved. The cross interference term in the intensity of an exciting and excited waves sum determining the hologram form includes several summands corresponding to various polarization components of electric field. If these summands have different spatial distributions in the hologram area and make comparable contributions into the hologram refractive index distribution, the hologram efficiency may decrease. But if one of the electric field polarization components of either exciting wave or a wave being excited is dominant over the others, the hologram configuration accounts phase relations between the exciting and the excited waves more correctly than simple gratings (3).

In addition to holograms of the type (2), we proposed and analyzed holograms recorded in two successive iterations in the same area in the PMMA layer. Such holograms are considerably more efficient than simple periodic gratings (3) and holograms (2) recorded in one iteration for all incidence angles of the exciting wave.

In our future publications, we will study in more details discrete volume holograms where refractive index can have only two values one of which is 1. Such discrete holograms are expected to be easier to fabricate (e.g. by 3D printing [29]) than the continuous holograms.

The proposed holographic method is universal, it can potentially be applied for excitation of not only plasmonic surface waves, but of other types of optical surface waves (e.g. Tamm waves in dielectric structures [30]) as well, and also for excitation of dielectric and plasmonic waveguide modes.

**Funding Information:** Advanced Research Foundation (7/004/2013-2018).

**REFERENCES**

1. A. Angelini, *Photon management assisted by surface waves on photonic crystal* (Springer International Publishing, Cham, Switzerland, 2017).
2. S. Pirotta, X. G. Xu, A. Delfan, S. Mysore, et al., "Surface-enhanced Raman scattering in purely dielectric structures via Bloch surface waves," J. Phys. Chem. C **117**, 6821-6825 (2013).
3. I. A. Nechepurenko, A. V. Dorofeenko, A. P. Vinogradov, I. N. Kurochkin, "Enhancement of Raman scattering by surface wave in photonic crystal," Moscow Univ. Chem. Bulletin **70**, 117-120 (2015).
4. J. Homola, "Surface plasmon resonance sensors for detection of chemical and biological species," Chem. Rev. **108**, 462-493 (2008).
5. N. Kinsey, M. Ferrera, V. M. Shalaev, A. Boltasseva, "Examining nanophotonics for integrated hybrid systems: a review of plasmonic interconnects and modulators using traditional and alternative materials," JOSA B **32**, 121-142 (2015).
6. S. Sun, A.-H. A. Badawy, V. Narayana, T. El-Ghazawi, V. J. Sorger, "The case for hybrid photonic plasmonic interconnects (HyPPIs): low-latency energy-and-area-efficient on-chip interconnects," IEEE Photonics Journal **7**, 4801614 (2015).
7. V. E. Babicheva, R. Malureanu, A. V. Lavrinenko, "Plasmonic finite-thickness metal-semiconductor-metal waveguide as ultra-compact modulator," Photonics and Nanostructures **11**, 323-334 (2013).
8. J. Grandidier, G. Colas des Francs, S. Massenot, A. Bouhelier, et al., "Leakage radiation microscopy of surface plasmon coupled emission: investigation of gain-assisted propagation in an integrated plasmonic waveguide," Journal of Microscopy **239**, 167-172 (2010).
9. B. Hecht, H. Bielefeldt, L. Novotny, Y. Inouye, and D. W. Pohl, "Local excitation, scattering, and interference of surface plasmons," Phys. Rev. Lett. **77**, 1889-1892 (1996).
10. M. G. Nielsen, J.-C. Weeber, K. Hassan, J. Fatome, et al., "Grating couplers for fiber-to-fiber characterizations of stand-alone dielectric loaded surface plasmon waveguide components," J. Lightwave Technol. **30**, 3118-3125 (2012).
11. A. I. Ignatov, A. M. Merzlikin, "Excitation of plasmonic waves in metal-dielectric structures by a laser beam using holography principles," Optics Communications (submitted).
12. A. Hessel, A. A. Oliner, "A new theory of Wood's anomalies on optical gratings," Appl. Optics **4**, 1275-1297 (1965).
13. D. Maystre, "Theory of Wood's anomalies," in *Plasmonics*, S. Enoch, and N. Bonod, eds. (Springer-Verlag, Berlin Heidelberg, 2012), pp. 39-83.
14. A. I. Ignatov, A. M. Merzlikin, and A. V. Baryshev, "Wood anomalies for s-polarized light incident on a one-dimensional metal grating and coupling of them with channel plasmons," Phys. Rev. A **95**, 053843 (2017).
15. I. Epstein, Y. Lilach, A. Arie, "Shaping plasmonic light beams with near-field plasmonic holograms," J. Opt. Soc. Am. B **31**, 1642-1647 (2014).
16. O. Avayu, I. Epstein, E. Eizner, T Ellenbogen, "Polarization controlled coupling and shaping of surface plasmon polaritons by nanoantenna arrays," Optics Letters **40**, 1520-1523 (2015).
17. Q. Xu, X. Zhang, Y. Xu, C. Ouyang, et al., "Polarization-controlled surface plasmon holography," Laser Photonics Rev. **11**, 1-9 (2016).
18. I. Epstein, Y. Tsur, A. Arie, "Surface-plasmon wavefront and spectral shaping by near-field holography," Laser Photonics Rev. **10**, 360-381 (2016).
19. J. J. Cowan, "Holography with standing surface plasma waves," Opt. Commun. **12**, 373-378 (1974).
20. J. Chen, L. Li, T. Li, S. N. Zhu, "Indefinite plasmonic beam engineering by in-plane holography," Scientific Reports **6**, 28926 (2016).
21. Y. Tsur, I. Epstein, R. Remez, A. Arie, "Wavefront shaping of plasmonic beams by selective coupling," ACS Photonics **4**, 1339-1343 (2017).
22. A. Minovich, A. E. Klein, N. Janunts, T. Pertsch, D. N. Neshev, Y. S. Kivshar, "Generation and near-field imaging of Airy surface plasmons," Phys. Rev. Lett. **107**, 116802 (2011).
23. O. You, B. Bai, X. Wu, Z. Zhu, Q. Wang, "A simple method for generating unidirectional surface plasmon polariton beams with arbitrary profiles," Optics Letters **40**, 5486-5489 (2015).
24. P. Genevet, J. Lin, M. A. Kats, F. Capasso, "Holographic detection of the orbital angular momentum of light with plasmonic photodiodes," Nature Commun. **3**, 1278 (2012).
25. P. Genevet, F. Capasso, "Holographic optical metasurfaces: a review of current progress," Rep. Prog. Phys. **78**, 024401 (2015).
26. J. Lin, J. P. B. Mueller, Q. Wang, G. Yuan, N. Antoniou, X.-C. Yuan, F. Capasso, "Polarization-controlled tunable directional coupling of surface plasmon polaritons," Science **340**, 331-334 (2013).

27. G. Beadie, M. Brindza, R. A. Flynn, A. Rosenberg, J. S. Shirk, "Refractive index measurements of poly(methyl methacrylate) (PMMA) from 0.4-1.6 µm," Appl. Opt. **54**, F139-F143 (2015).
28. P. B. Johnson, R. W. Christy, "Optical constants of the noble metals," Phys. Rev. B **6**, 4370-4379 (1972).
29. M. Malinauskas, M. Farsari, A. Piskarskas, S. Juodkazis, "Ultrafast laser nanostructuring of photopolymers: a decade of advances," Physics Reports **533**, 1-31 (2013).
30. A. P. Vinogradov, A. V. Dorofeenko, A. M. Merzlikin, A. A. Lisyansky, "Surface states in photonic crystals," Physics-Uspekhi **53**, 243-256 (2010).